\documentclass[pra,showpacs,amsmath,amssymb, twocolumn,10pt]{revtex4}

\usepackage{amsthm}
\usepackage{dcolumn}
\usepackage{bm}
\usepackage{graphicx}

\begin{document}

\newtheorem{corollary}{Corollary}
\newtheorem{definition}{Definition}
\newtheorem{example}{Example}
\newtheorem{lemma}{Lemma}
\newtheorem{proposition}{Proposition}
\newtheorem{conjecture}{Conjecure}
\newtheorem{theorem}{Theorem}
\newtheorem{fact}{Fact}
\newtheorem{property}{Property}
\newcommand{\bra}[1]{\langle #1|}
\newcommand{\ket}[1]{|#1\rangle}
\newcommand{\braket}[3]{\langle #1|#2|#3\rangle}
%%inner product
\newcommand{\ip}[2]{\langle #1|#2\rangle}
%%outer product
\newcommand{\op}[2]{|#1\rangle \langle #2|}

\newcommand{\tr}{{\rm Tr}}
\newcommand {\cE } {{\mathcal{E}}}
\newcommand {\cF } {{\mathcal{F}}}
\newcommand {\cM } {{\mathcal{M}}}
\newcommand {\cB } {{\mathcal{B}}}
\newcommand {\cV } {{\mathcal{V}}}
\newcommand {\cU } {{\mathcal{U}}}
\newcommand {\diag } {{\rm diag}}
\newcommand{\cH}{\mathcal{H}}
\newcommand{\cK}{\mathcal{K}}
\newcommand{\cZ}{\mathcal{Z}}

\newcommand{\rL}{\mathrm{L}}
\newcommand{\rT}{\mathrm{T}}
\newcommand{\rU}{\mathrm{U}}
\newcommand{\rM}{\mathrm{M}}
\newcommand{\rS}{\mathrm{S}}
\newcommand {\rJ} {{\mathrm{J}}}
\newcommand{\rA}{\mathrm{A}}
\newcommand{\rC}{\mathrm{C}}

\newcommand {\bU } {{\mathbb{U}}}
\newcommand {\bS } {{\mathbb{S}}}
\newcommand {\bL } {{\mathbb{L}}}

%%%%%%%%%%%%%%%%%%%%%%%%%%%%%%%%%%
\newcommand{\real}{{\mathbb R}}
\newcommand{\ent}{{\mathbb Z}}
\newcommand{\com}{{\mathbb C}}
\newcommand{\un}{1\mkern -4mu{\textrm l}}
\newcommand{\FF}{{\mathbb F}}
\newcommand{\T}{{\mathbb T}}

\newcommand{\A}{{\mathcal A}}
\newcommand{\B}{{\mathcal B}}
\newcommand{\E}{{\mathcal E}}
\newcommand{\F}{{\mathcal F}}
\newcommand{\G}{{\mathcal G}}
\renewcommand{\H}{{\mathcal H}}
\renewcommand{\L}{{\mathcal L}}
\newcommand{\M}{{\mathcal M}}
\newcommand{\N}{{\mathcal N}}
\renewcommand{\P}{{\mathcal P}}
\newcommand{\R}{{\mathcal R}}
\renewcommand{\S}{{\mathcal S}}
\newcommand{\U}{{\mathcal U}}

\renewcommand{\a}{\alpha}
\renewcommand{\b}{\beta}
\newcommand{\g}{\gamma}
\newcommand{\Ga}{\Gamma}
\renewcommand{\d}{\delta}
\newcommand{\D}{\Delta}
\newcommand{\e}{\varepsilon}
\renewcommand{\l}{\lambda}
\newcommand{\La}{\Lambda}
\renewcommand{\e}{\varepsilon}
\newcommand{\f}{\varphi}
\renewcommand{\O}{{\Omega}}
\renewcommand{\o}{{\omega}}
\newcommand{\s}{\sigma}
\renewcommand{\t}{\theta}
\newcommand{\ma}{{\mathbb M}}

\newcommand{\Tr}{\mbox{\rm Tr}}
\newcommand{\ot}{\otimes}
\newcommand{\8}{\infty}
\newcommand{\el}{\ell}
\newcommand{\pa}{\partial}
\newcommand{\la}{\langle}
\newcommand{\ra}{\rangle}
\newcommand{\wt}{\widetilde}
\newcommand{\wh}{\widehat}
\newcommand{\n}{\noindent}
\newcommand{\pf}{\noindent{\it Proof.~~}}
\newcommand{\cqd}{\hfill$\Box$}
\newcommand{\be}{\begin{eqnarray*}}
\newcommand{\ee}{\end{eqnarray*}}
\newcommand{\beq}{\begin{equation}}
\newcommand{\eeq}{\end{equation}}
\newcommand{\dis}{\displaystyle}
%%%%%%%%%%%%%%%%%%%%%%%%

\title{Bounds on the distance between a unital quantum channel and the convex hull of unitary channels, with applications to the asymptotic quantum Birkhoff conjecture}

\author{Nengkun Yu}
\email{nengkunyu@gmail.com}
\affiliation{State Key Laboratory of Intelligent Technology and Systems, Tsinghua National Laboratory for Information Science and Technology, Department of Computer Science and Technology, Tsinghua University, Beijing 100084, China\\
and Centre for Quantum Computation and Intelligent Systems (QCIS), Faculty of
Engineering and Information Technology, University of Technology,
Sydney, NSW 2007, Australia}

\author{Runyao Duan}
\email{runyao.duan@uts.edu.au}
\affiliation{Centre for Quantum Computation and Intelligent Systems (QCIS), Faculty of
Engineering and Information Technology, University of Technology,
Sydney, NSW 2007, Australia\\
and State Key Laboratory of Intelligent Technology and Systems, Tsinghua National Laboratory for Information Science and Technology, Department of Computer Science and Technology, Tsinghua University, Beijing 100084, China
}

\author {Quanhua Xu}
\email{qxu@univ-fcomte.fr}
\affiliation{School of Mathematics and Statistics, Wuhan
    University, Wuhan 430072, China\\
    and Laboratoire de Math\'ematiques, Universit\'e de Franche-Comt\'e, 25030 Besan\c
    con cedex, France }
\date{December 28, 2011}

\begin{abstract}
Motivated by the recent resolution of Asymptotic Quantum Birkhoff Conjecture (AQBC), we attempt to estimate the distance between a given unital quantum channel and the convex hull of unitary channels. We provide two lower bounds on this distance by employing techniques from quantum information and operator algebras, respectively. We then show how to apply these results to construct some explicit counterexamples to AQBC. We also point out an interesting connection between the Grothendieck's inequality and AQBC.
\end{abstract}


\maketitle

\section{Introduction}
Suppose we are given a quantum system with a $d$-dimensional Hilbert space $\cH_d$, and the state (or density operator) of the system is given by a trace one positive operator $\rho$ from the linear operator space $\rL(\cH_d)$. Quantum channels, or trace-preserving completely positive maps, are all possible deterministic quantum operations one can perform over the system \cite{Kra83,Cho75}. Let $\Phi$ be such a quantum channel over $\rL(\cH_d)$ with Kraus operator sum representation $\Phi=\sum_k E_k\cdot E_k^\dagger$, and let $K(\Phi)={\rm span}\{E_k\}$ be its Kraus operator space. The convex hull of unitary channels (noiseless channels) on $\rL(\cH_d)$ is given by ${\rm Conv}(\bU(\cH_d))$. So any $\Psi\in {\rm Conv}(\bU(\cH_d))$ can be written as a mixture (convex combination) of unitary channels. (The number of unitary channels in the mixture can be made finite due to the Carath$\acute{\rm e}$odory's theorem on convex hull). The mixture of unitary channels plays a special role in environment-assisted quantum communication model. Actually, these channels can be made noiseless for quantum information transmission with the help of a friendly environment even in one-shot case. Furthermore, it turns out that these channels are the only quantum channels having this desirable property \cite{GW02}. Surprisingly, if arbitrarily large number of uses of the channels are allowed, unital quantum channels, those channels $\Phi$ with identity operator a fixed point, say $\Phi(I)=I$, can also achieve maximum capacity and act exactly like noiseless channel \cite{SVW05}.

Clearly, any mixture of unitary channels remains unital. An interesting question is to ask whether one can reverse this procedure, i.e., decomposing any unital quantum channel $\Phi\in \rT(\cH_d)$ into a mixture of unitary channels from $\bU(\cH_d)$. This was called ``quantum Birkhoff conjecture" (QBC), originated from  Birkhoff's celebrated characterization of the extreme points of doubly stochastic matrices. Unfortunately, this conjecture is only true for $d\leq 2$, and counterexamples exist whenever $d\geq 3$ \cite{KB86,KH87,LS93}. This suggests the following quantity to measure the distance between $\Phi$ and the convex hull of unitary channels.
$$D(\Phi, {\rm Conv}(\bU(\cH)))=\inf\{D(\Phi, \Psi):\Psi\in {\rm Conv}(\bU(\cH))\},$$
where $D(\Phi,\Psi)$ will be given by the diamond norm of $\Phi-\Psi$. Since ${\rm Conv}(\bU(\cH))$ is a compact convex set, ``${\rm inf}$" in the above equation can be replaced by ``${\rm min}$".

Motivated by some results in about the environment-assisted quantum capacity and in an attempt to remedy the conjecture in certain way, Smolin, Verstraete, and Winter proposed the following
\begin{conjecture}(Asymptotic Quantum Birkhoff Conjecture \cite{SVW05}) Let $\Phi\in \rT(\cH)$ be a unital channel, then $\Phi^{\otimes n}$ can be approximated by a mixture of unitary channels from $\bU(\cH^{\otimes n})$ with arbitrary precision. That is
$${\rm lim}_{n\rightarrow\infty}D(\Phi^{\otimes n}, {\rm Conv}(\bU(\cH^{\otimes n})))=0.$$
\end{conjecture}
This revised conjecture seems highly reasonable as one could naturally expect that many copies of a unital channel will be better approximated by a mixture of unitary channels on a higher-dimensional space. If this is true, it will provide a very satisfactory interpretation to the following result: The environment-assisted quantum capacity of any unital channel over $\rL(\cH_d)$ is given by $\log_2 d$ qubits, the maximum capacity one can achieve under this model. A much more deep consequence is that the structure of unital channels will be greatly simplified. Due to its significance, the asymptotic quantum Birkhoff conjecture was listed as one of major open problems in quantum information theory \cite{Werner05}.

Some supporting evidences were obtained in Ref. \cite{MW08}, where Mendl and Wolf presented a unital channel $\Phi$ such that $\Phi^{\otimes 2}$ is a mixture of unitary channels although $\Phi$ itself is not. Furthermore, they showed that it is possible that the tensor of $\Phi$ and a constant unital channel (a completely depolarizing channel that maps every state into the completely mixed state $I/d$) may become a mixture of unitary channels. One may naturally conjecture these properties might be true for any unital quantum channels.

Recently Haagerup and Musat disproved this asymptotic version by exhibiting a class of so-called non-factorizable maps as counterexamples \cite{HM2011a}. Actually the results obtained in Ref. \cite{HM2011a} shows that any such non-factorizable map $\Phi$ is a very strong counterexample to AQBC in the following sense:
$$D(\Phi\otimes \Psi, {\cF\cM} (\rL(\cH_d\otimes\cH_m))\geq D(\Phi, {\cF\cM}(\rL(\cH_d))),$$
where $\Psi$ is any unital channel over $\rL(\cH_m)$, and $\cF\cM(\rL(\cH_d))$ denotes the set of factorizable maps over $\rL(\cH_d)$. In other words, any non-factorizable map tensoring with a unital channel could not reduce the distance to the set of factorizable maps, which is a super-set of the convex hull of unitary channels. See also Shor's talk in Ref. \cite {Shor2010} for an alternative approach to AQBC and an excellent discussion of the results in Ref. \cite{HM2011a}. The interesting thing here is that all these counterexamples are non-factorizable maps, and it remained unknown whether any facterizable map would fulfill AQBC. This problem was signified in the arXiv version of Ref. \cite{HM2011a} by establishing the following surprising connection: If all factorizable maps satisfy AQBC, then the Connes embedding problem has a positive answer.

Motivated by these progresses and in order to better understand the structure of unital channels, in this paper we are interested in estimating the trace distance between a unital quantum channel and the convex hull of unitary channels, say $D(\Phi, {\rm Conv}(\bU(\cH)))$. We find that this distance is interesting even from the perspective of quantum channel discrimination: Suppose we are given an unknown quantum channel, which is secretly chosen between $\Phi$ and some $\Psi\in {\rm Conv}(\bU(\cH)))$ with equal probability $1/2$. Then due to the operational meaning of trace distance, we can conclude that the success probability of discrimination is at least $1/2+1/4D(\Phi, {\rm Conv}(\bU(\cH)))$, which is strictly larger than $1/2$ whenever $\Phi$ is not a mixture of unitary channels. Another purpose of this paper is to provide some relatively elementary and self-contained disproofs to AQBC. This is partially due to the fact that the elegant disproof of AQBC in Ref. \cite{HM2011a} makes use of some basic properties of factorizable maps which cannot be easily appreciated by readers who do not have deep background in operator algebras.

In Section II we collect some preliminaries about super-operators and Schur channels. Then in Section III we explain in detail the operational meaning of trace distance. In Section IV we first provide a computable lower bound for $D(\Phi, {\rm Conv}(\bU(\cH)))$ when the Kraus operator space of $\Phi$ does not contain any unitary operator. This enables us to derive many counterexamples for AQBC, including some factorizable maps presented in Ref. \cite{HM2011a}. It is worth pointing out that this proof only employs some basic techniques from quantum information theory. We believe that it may interest readers with quantum information background. In Section V we go further to study the class of Schur channels. In this special case, we are able to provide a lower bound and an upper bound for $D(\Phi, {\rm Conv}(\bU(\cH)))$. Roughly speaking, we show that up to a factor of $1/2$, any Schur channel can be approximated by a mixture of diagonal unitary channels, and the later has a simpler structure. As a direct application, we obtain a new proof of the fact that any Schur channel that does not satisfy the QBC will automatically violate the AQBC. Our proof for this part has employed some powerful tools from operator algebras. In Section VI we present two explicit examples of Schur channels to demonstrate the utility of our results: the first example has only two Kraus operators and is a non-factorizable map, and the second one is a factorizable map. As another interesting application, in Section VII we point out a connection between AQBC and Grothendieck's inequality in the metric theory of tensor products.

\noindent {\bf Remarks on related results:} After we obtained the results in Section IV, and were working on the proof of the Theorem \ref{schur-multiplier1} in Section V, the second author R.D. happened to learn from Prof. M. B. Ruskai that Haagerup and Musat had made further progress on the connection between Schur channels and AQBC. Namely, they obtained Theorem \ref{hm-theorem1} and thus showed that any Schur channel that violates QBC (including some factorizable maps) should also be a counterexample to AQBC \cite{HM2011b}. They also provided a modified version of the connection between factorizable maps satisfying AQBC and Connes embedding problem. The proof of Theorem \ref{schur-multiplier1} has employed some similar techniques in \cite{HM2011b}.

\section{Preliminaries}
We will use symbols $\cH$, $\cK$, etc to represent finite dimensional Hilbert spaces over complex numbers. A $d$-dimensional Hilbert space $\cH$, which is essentially the same as $\mathcal{C}^{d}$, will be explicitly represented as $\cH_d$ whenever the dimension matters. $\rL(\cH,\cK)$ denotes the set of linear operators (or mappings) from $\cH$ to $\cK$, and $\rL(\cH)$ is shorthand for $\rL(\cH,\cH)$. For any $X\in L(\cH)$, $X^\dagger\in L(\cH)$ denotes the adjoint operator (or complex conjugate) of $X$. $X$ is Hermitian (or self-adjoint) if $X^\dagger=X$. $U\in \rL(\cH)$ is said to be unitary if $U^\dagger U=I_\cH$. We denote the set of unitary operators on $\cH$ by $\rU(\cH)$. $X\in \rL(\cH)$ is (semi-definite) positive, write $X\geq 0$, if the quadratic form $\braket{\psi}{X}{\psi}\geq 0$ for any $\ket{\psi}\in \cH$. In particular, $X$ is said to be a density operator (or a quantum state) if $X$ is positive and with trace one. $\rT(\cH,\cK)$ is the set of linear mappings
 from $\rL(\cH)$ to $\rL(\cK)$. Again, $\rT(\cH)$ is shorthand for $\rT(\cH,\cH)$. Elements in $\rT(\cH,\cK)$ are normally called super-operators. Note that $\rL(\cH)$ is a Hilbert space with the standard Hilbert-Schmidt inner product $<A,B>=\tr(A^\dagger B)$. Then the adjoint operator of $\Phi\in \rT(\cH,\cK)$ is defined as the unique super-operator $\Phi^\dagger\in \rT(\cK,\cH)$ such that
 $$<Y,\Phi(X)>=<\Phi^\dagger(Y),X>,~~\forall X\in\rL(\cH),Y\in\rL(\cK).$$
A super-operator $\Phi\in \rT(\cH,\cK)$ is said to be positive if it preserves the positivity, say, $\Phi(X)\geq 0$ whenever $X\geq 0$. $\Phi$ is said to be a quantum channel if it satisfies: i) (trace-preserving) $\tr(\Phi(X))=\tr(X)$ for any $X\in \rL(\cH)$, and ii)(completely positive) for any $n\geq 1$, the induced super-operator $\Phi_n=\Phi\otimes I_{\rL(\cH_n)}\in \rT(\cH\otimes \cH_n, \cK\otimes \cH_n)$ is positive, where ${I}_{\rL(\cH_n)}$ is the identity super-operator on $\rL(\cH_n)$. We call $\Phi$ a quantum unital channel if it further satisfies: iii) (unital condition) $\Phi(I_\cH)=I_\cK$. Any unitary operator $U\in\rU(\cH)$ induces a unitary quantum channel $\cU\in \rT(\cH)$ in the following way: $\cU(X)=UXU^\dagger$. The class of unitary channels on $\rL(\cH)$ will be denoted as $\bU(\cH)$.

Any super-operator $\Phi\in \rT(\cH,\cK)$ can be represented by a pair of linear operators $A,B\in \rL(\cH,\cK\otimes \cZ)$ such that
\begin{equation}\label{stinespring-1}
\Phi(X)=\tr_{\cZ}AXB^\dagger,~~X\in\rL(\cH),
\end{equation}
where $\cZ$ is an auxiliary Hilbert space with ${\rm dim}(\cZ)\leq {\rm dim}(\cH){\rm dim}(\cK)$, and $\tr_{\cZ}$ represents the partial trace over $\cZ$. For the special case of quantum channels, the above form can be greatly simplified. Actually, in Eq. (\ref{stinespring-1}) we can choose $A=B=V\in \rL(\cH,\cK\otimes \cZ)$ for some isometry $V$ and obtain the following well-known Stinespring unitary embedding representation of a quantum channel:
\begin{equation}\label{stinespring-2}
\Phi(X)=\tr_{\cZ}VXV^\dagger,~~V^\dagger V=I_\cH.
\end{equation}

If we specify an orthonormal basis $\{\ket{k_\cZ}\}$ of $\cZ$, we can rewrite $\Phi$ in Eq (\ref{stinespring-1}) into the following form:
\begin{equation}\label{kraus-1}
\Phi(X)=\sum_{k=1}^{\rm dim(\cZ)} A_k X B_k^\dagger,
\end{equation}
where $A_k=\bra{k_\cZ}A$ and $B_k=\bra{k_\cZ}B$ are linear operators in $\rL(\cH,\cK)$.
Similarly, when $\Phi$ is a quantum channel, we can choose $A_k=B_k=\bra{k_\cZ}V$ so that
\begin{equation}\label{kraus-2}
\Phi(X)=\sum_{k=1}^{\rm dim(\cZ)} A_k X A_k^\dagger,~~\sum_k A_k^\dagger A_k=I_\cH,
\end{equation}
which is the famous Kraus operator sum representation of a quantum channel \cite{Kra83}.

Now we tend to introduce norms of super-operators in $\rT(\cH,\cK)$ based on the norms of linear operators. We refer to Refs. \cite{Watrous04,Watrous09} for some detailed discussion on norms of super-operators and how to compute them using semi-definite programming techniques. We will briefly review some basic results for later use. For any $X\in \rL(\cH_d)$ and $p\geq 1$, the $p$-th norm of $X$ is given by
$$||X||_p=(\tr|X|^p)^{\frac{1}{p}},$$
where $|X|=\sqrt{X^\dagger X}.$
The trace and the operator norms of $X$ are special cases of $p=1$ and $p\rightarrow \infty$, respectively,
$$||X||_1=\tr|X|,~||X||_\infty=\max_{\ip{\psi}{\psi}=1}||X\ket{\psi}||.$$

The trace norm and the operator norm of a super-operator $\Phi\in \rT(\cH,\cK)$ are given respectively as follows:
$$||\Phi||_1=\sup_{||X||_1\leq 1}||\Phi(X)||_1,~~||\Phi||_\infty=\sup_{||X||_\infty\leq 1}||\Phi(X)||_\infty.$$ In the above equation we can replace``$\sup$"  with ``$\max$" when only finite dimensional Hilbert spaces are involved. The completely bounded trace norm (or diamond norm) and operator norm (simply completely bounded norm) are given respectively as follows:
$$||\Phi||_\diamond=\sup_{n\geq 1}||\Phi\otimes I_{\rL(\cH_n)}||_1,~~||\Phi||_{\rm cb}=\sup_{n\geq1}||\Phi\otimes I_{\rL(\cH_n)}||_{\infty}.$$

\begin{proposition} For any $\Phi\in \rT(\cH,\cK)$, the diamond norm and the completely bounded norm satisfy the following properties:
\begin{itemize}
\item i)  The dimension of the auxiliary system to achieve the norms can be restricted to that of $\cH$,$||\Phi||_\diamond=||\Phi\otimes I_{\rL(\cH})||_1$ and $||\Phi||_{\rm cb}=||\Phi\otimes I_{\rL(\cH)}||_{\infty}.$
\item ii) The following duality relation holds for $\Phi$ and $\Phi^\dagger$,
$||\Phi||_1=||\Phi^\dagger||_\infty$ and $||\Phi||_\diamond=||\Phi^\dagger||_{\rm cb}.$

\item iii) If $\Phi$ is completely positive, then
$||\Phi||_\diamond=||\Phi||_1$ and $||\Phi||_{\rm cb}=||\Phi||_\infty=||\Phi(I_{\cH})||_\infty.$
\end{itemize}
\end{proposition}

The norms defined above enable us to introduce distance between quantum states and quantum channels. The trace distance between two quantum density operators $\rho$ and $\sigma$ in $\rL(\cH)$ is given by
$$D(\rho,\sigma)={||\rho-\sigma||_1}.$$
In the following discussion we also need the fidelity between $\rho$ and $\sigma$,
$$F(\rho,\sigma)=\tr \sqrt{\rho^{1/2}\sigma \rho^{1/2}}.$$ The so-called Uhlmann theorem makes the meaning of fidelity more transparent:
$$F(\rho,\sigma)=\max_{\ket{\psi},\ket{\phi}}|\ip{\psi}{\phi}|,$$
where $\ket{\psi},\ket{\phi}\in \cH\otimes \cK$ range over all purifications of $\rho$ and $\sigma$, respectively, say  $\tr_\cK \op{\psi}{\psi}=\rho$ and $\tr_\cK \op{\phi}{\phi}=\sigma$. Most notably, the above equation remains true even when one of $\ket{\psi}$ or $\ket{\phi}$ is fixed.  This fact plays a crucial role in our later discussion. Trace distance and fidelity are equivalent in characterizing the distance between two states in the following sense:
 $$2(1-F(\rho,\sigma))\leq D(\rho,\sigma)\leq 2\sqrt{1-F^2(\rho,\sigma)}.$$

Following the same idea, we can define the trace distance between two quantum channels $\Phi$ and $\Psi$ via the following way:
 $$D(\Phi,\Psi)={||\Phi-\Psi||_{\diamond}}.$$

Let us now introduce a special class of super-operators. For any $S\in \rL(\cH_d)$, we can define a super-operator $\Phi_S$ via the following way:
 $$\Phi_S(X)=S\circ X, ~\forall X\in \rL(\cH_d),$$
 where $S\circ X=[s_{kj}x_{kj}]$ is the entry-wise product or Hadamard product. (Here we assume that we have specified an orthonormal basis $\{\ket{k}:k=1,\cdots,d\}$ for $\cH_d$. Thus any linear operator from $\rL(\cH_d)$ is expressed as a matrix under the standard matrix basis $\{\op{k}{j}\}$. For instance, $S=\sum_{k,j}s_{kj}\op{k}{j}$). Such $\Phi_S$ is called Schur multiplier induced by $S$. Schur multipliers have been extensively studied in the literatures of operator algebras. We refer to Chapters $3$ and $8$ of Ref. \cite{Paulsen02} for some highly accessible introductions, and Ref. \cite{HM2011a} for recent advances. For later use, some basic properties of Schur multipliers are listed as follows:
\begin{proposition}\label{schur0}
Let $S\in \rL(\cH_d)$. Then $\Phi_S$ satisfies the following:
\begin{itemize}
\item i) $\Phi_S=\sum_{k=1}^d A_k \cdot B_k^\dagger$, where all $A_k,B_k$ are diagonal matrices;
\item ii) $\Phi_S$ is positivity-preserving iff $S$ is positive;
\item iii) $\Phi_S$ is completely positive iff $S$ is positive;
\item iv) $\Phi_S$ is trace-preserving if $s_{kk}=1$ for $k=1,\cdots,d$;
\item v) $\Phi_S$ is unital iff $s_{kk}=1$ for $k=1,\cdots,d$.
\end{itemize}
\end{proposition}

{\bf Proof:} $iv)$ and $v)$ follow directly by evaluating $\tr(\Phi_S(\op{k}{j}))=s_{kj}\delta_{kj}$. We shall see that $ii)$ and $iii)$ are simple corollaries of $i)$. So we first prove $i)$. In fact, let $A$ and $B$ be any two $d\times d$ matrices such that $S=AB^\dagger$. We may assume $A=[a_1,\cdots, a_d]$ and $B=[b_1,\cdots, b_d]$, where $a_k$ and $b_k$ are all $d$-dimensional column vectors. Then $S=\sum_k a_kb_k^\dagger$. Set $A_k={\rm Diag}(a_k)$, $B_k={\rm Diag}(b_k)$. That is, $A_k$ and $B_k$ are diagonal matrices with diagonals $a_k$ and $b_k$, respectively. By some routine calculations we directly verify that $\Phi_S(X)=\sum_{k=1}^d A_k X B_k^\dagger$ for any $X\in \rL(\cH_d)$. In particular, when $S$ is positive we can write $S=AA^\dagger$ for some $A\in \rL(\cH_d)$. Hence we can choose $A_k=B_k$ in this special case. That proves both the positivity and completely positivity of $\Phi_S$. Conversely, if $\Phi_S$ is positive. Then by choosing $\ket{e}=\sum_{k=1}^d
 \ket{k}$, we have $\Phi_S(\op{e}{e})=S$ is positive. \hfill $\square$

The following proposition gives another fundamental property of Schur multiplier. Relevant discussions can be found in Page $110$ of Ref. \cite{Paulsen02}.
\begin{proposition}\label{schur-multiplier3}
For any Schur multiplier $\Phi$, the diamond norm, the trace norm, completely bounded norm and operator norm all coincide, that is, $||\Phi||_\diamond=||\Phi||_1=||\Phi||_{\rm cb}=||\Phi||_\infty.$
\end{proposition}

So a Schur multiplier $\Phi_S$ is a quantum channel iff $S$ is positive and with all diagonal entries one. In particular, whenever $\Phi_S$ is a quantum channel, it is also unital. We shall denote
$$\rS(\cH_d)=\{\Phi_S: S\in \rL(\cH_d), S\geq 0, s_{kk}=1,1\leq k\leq d\},$$
and call the elements from $\rS(\cH_d)$ (or simply $\rS_d$) Schur channels. Note that the difference of two Schur multipliers is still a Schur multiplier. Applying Proposition \ref{schur-multiplier3}, we obtain an immediate consequence that auxiliary systems are not required to distinguish between two Schur channels.

\section{Operational interpretation of trace distance}

We have introduced trace distance between quantum states and quantum channels, and will study the trace distance between a unital quantum channel and the convex hull of unitary channels in greater detail. Before we proceed, we need justify the importance of this measure from the perspective of quantum information. In one word, the trace distance characterizes some sort of stochastic distinguishability of quantum states and quantum channels. Actually, the trace distance naturally occurs when we study the following state discrimination problem. Suppose we are given an unknown quantum system whose state is secretly prepared in one of $\rho_0$ and $\rho_1$, with equal priori probability $1/2$. The task here is to determine the identity of the system with a success probability as high as possible. To do so we need apply a two-outcome quantum measurement $\{E_0,E_1\}$ to the system, and to maximize the success probability of discrimination, i.e.,
 $$P_{\rm succ}(\rho_0,\rho_1)=\max_{\{E_0,E_1\}}\frac{1}{2}(\tr\rho_0 E_0+\tr \rho_1 E_1),$$
where $E_i\geq 0$ and $E_0+E_1=I$. By some simple algebraic manipulations, one can verify that the optimal success probability of discrimination is given by \cite{Helstrom76}
 $$P_{\rm succ}(\rho_0,\rho_1)=\frac{1}{2}+\frac{1}{4}D(\rho_0,\rho_1).$$
Thus a larger trace distance between $\rho_0$ and $\rho_1$ implies a higher success probability of discrimination. This interpretation can be extended to compact convex sets of density operators. Let $\rA_0$ and $\rA_1$  be two compact convex sets of density operators. The trace distance between $\rA_0$ and $\rA_1$ is given by
$$D(\rA_0,\rA_1)=\min\{D(\rho_0,\rho_1):\rho_i\in \rA_i, i=0,1\}.$$
Then the optimal discrimination probability between $\rA_0$ and $\rA_1$ is given as
\begin{equation}\label{dis-sets}
P_{\rm succ}(\rA_0,\rA_1)=\frac{1}{2}+\frac{1}{4}D(\rA_0,\rA_1).
\end{equation}
The above formula indicates that we can operationally distinguish between two compact convex sets of density operators by performing a universal quantum measurement, and the success probability of discrimination is completely characterized by the trace distance between $\rA_0$ and $\rA_1$. The most surprising thing here is that the quantum measurement we perform does not depend on the exact form of the unknown state except the assumption that it is from one of $\rA_0$ and $\rA_1$.

It seems that Eq. (\ref{dis-sets}) was first obtained by Gutoski and Watrous in Ref. \cite{GW04} by using the convex set separation theorem. Jain  provided a different way based on the minimax theorem \cite{Jain05}. For completeness, we will outline the later approach as follows. Let $\{E_0,E_1\}$ be the quantum measurement we need perform, and $\rho_0$ and $\rho_1$ be two states from $\rA_0$ and $\rA_1$, respectively. Then the optimal success probability is given by
$$P_{\rm succ}(\rA_0,\rA_1)=\max_{\{E_i\}} \min_{\rho_i\in \rA_i}\frac{1}{2}(\tr \rho_0 E_0+\tr \rho_1 E_1).$$
The crucial point here is that we first take ``min'' over all possible pair of states $\rho_0$ and $\rho_1$ according to a fixed measurement $\{E_0,E_1\}$, and then take ``max" over all possible measurements to maximize the success probability of discrimination. Noticing that the objective function is linear in $(E_0,E_1)$ and $(\rho_0,\rho_1)$ when one of them is fixed, and all involving sets are compact convex, we can apply appropriate form of Sion's minimax theorem \cite{Sion58} to exchange the order of ``max'' and ``min'', and obtain Eq. (\ref{dis-sets}) immediately.

Now we try to generalize the above result to the case of quantum channels. The simplest case is to distinguish between two quantum channels $\Phi_0,\Phi_1\in \rT(\cH,\cK)$. The basic strategy here is to choose an input state $\rho\in \rL(\cH'\otimes \cH)$, and then to distinguish between the respective output states $I_{\rL(\cH')}\otimes \Phi_i (\rho)$, where $\cH'$ is a finite-dimensional auxiliary state space. We have
$$D(\Phi_0,\Phi_1;\rho)=D((I_{\rL(\cH')}\otimes \Phi_0) (\rho),(I_{\rL(\cH')}\otimes \Phi_1) (\rho)).$$
To achieve the maximum success probability, we need take ``sup" over all possible input states, and have
$$D(\Phi_0,\Phi_1)=\sup_{\rho}D(\Phi_0,\Phi_1;\rho).$$
One can readily verify that the RHS of the above equation gives us the diamond norm $||\Phi_0-\Phi_1||_\diamond$, and $\rho$ can be restricted to density operators on $\cH\otimes \cH$ (thus ``sup" can be replaced as ``max"). To generalize the trace distance to compact convex sets of quantum channels, we first need the trace distance with input state $\rho$ as follows:
$$\widetilde{D}(\rC_0,\rC_1;\rho)=\min_{\Phi_i\in \rC_i}D(\Phi_0,\Phi_1;\rho).$$
Then the final resulting operational trace distance between $\rC_0$ and $\rC_1$ is given by
\begin{equation}\label{tildeD}
\widetilde{D}(\rC_0,\rC_1)=\sup_{\rho}\widetilde{D}(\rC_0,\rC_1;\rho)=\sup_{\rho}\min_{\Phi_i\in \rC_i}D(\Phi_0,\Phi_1;\rho),
\end{equation}
where $\rho$ ranges over all possible bipartite density operators on $\cH'\otimes \cH$, and it is not clear whether we can replace ``sup" with ``max" as the dimension of $\cH'$ may be arbitrarily large. The optimal success probability of discrimination between $\rC_0$ and $\rC_1$ is given by
$$P_{\rm succ}(\rC_0,\rC_1)=\frac{1}{2}+\frac{1}{4}\widetilde{D}(\rC_0,\rC_1).$$
Interestingly, the (ordinary) trace distance between $\rC_0$ and $\rC_1$ is given by
$$D(\rC_0,\rC_1)=\min_{\Phi_i\in \rC_i}D(\Phi_0,\Phi_1)=\min_{\Phi_i\in \rC_i}\max_{\rho}D(\Phi_0,\Phi_1;\rho).$$
The major difference between $D(\rC_0,\rC_1)$ and $\widetilde{D}(\rC_0,\rC_1)$ is that the orders of ``max" (``sup") and ``min" has been reversed. It is not obvious that whether the orders of ``min" and ``max" (``sup") are exchangeable or not as it is unclear whether the objective function $D(\Phi_0,\Phi_1;\rho)$ satisfies the requirements of minimax theorem. Consequently, it seems not clear whether $\widetilde{D}(\rC_0,\rC_1)$ is the same as  $D(\rC_0,\rC_1)$. Nevertheless, we still have
$$\widetilde{D}(\rC_0,\rC_1;\rho)\leq \widetilde{D}(\rC_0,\rC_1)\leq D(\rC_0,\rC_1).$$ In particular, we have the following simple property.
\begin{property}
Let $\rC_0, \rC_1\subseteq \rT(\cH,\cK)$ be two compact convex sets of quantum channels, and let $\rho$ be a bipartite pure entangled state over $\cH\otimes \cH$ with full Schmidt rank. Then the following are equivalent:
\begin{itemize}
\item[i).] $\rC_0\cap \rC_1=\emptyset$;
\item[ii).] $D(\rC_0,\rC_1)>0$;
\item[iii).] $\widetilde{D}(\rC_0,\rC_1)>0$; and
\item[iv).] $\widetilde{D}(\rC_0,\rC_1;\rho)>0$.
\end{itemize}
\end{property}

{\bf Proof:} We only need to establish the equivalence between i) and iv). By definition, iv) means that we can distinguish between $\rC_0$ and $\rC_1$ using $\rho$ as an input. This immediately implies that $C_0$ and $C_1$ should be disjoint. In other words, i) should hold. The direction that i)$\Rightarrow$ iv) is a little bit tricky, and the key here is to apply a generalized form of Choi isomorphism \cite{Cho75} between super-operators and bipartite linear operators. By contradiction, assume that $\rC_0$ and $\rC_1$ are disjoint but $\widetilde{D}(\rC_0,\rC_1;\rho)=0$. It follows from the definition that there exist $\Phi_0\in \rC_0$ and $\Phi_1\in \rC_1$ such that
$$
D((I_{\rL(\cH)}\otimes \Phi_0) (\rho),(I_{\rL(\cH)}\otimes \Phi_1) (\rho))=0.
$$
Equivalently, we have
\begin{equation}\label{choi-ge}
(I_{\rL(\cH)}\otimes \Phi_0) (\rho)=(I_{\rL(\cH)}\otimes \Phi_1) (\rho).
\end{equation}
Noticing that $\rho$ is a bipartite pure state with full Schmidt rank, we have the following generalized Choi-isomorphism:
$$\rJ: \Phi\mapsto (I_{\rL(\cH)}\otimes \Phi) (\rho).$$
(The standard Choi-isomorphism is to choose $\rho$ as the maximally entangled state $\ket{\Omega}=1/\sqrt{d}\sum_{k=1}^d \ket{k}\ket{k}$). Applying this isomorphism, we deduce from Eq. (\ref{choi-ge}) that $\Phi_0=\Phi_1$. This contradicts the assumption $\rC_0\cap \rC_1=\emptyset$. \hfill $\square$

So whenever two compact convex sets of quantum channels are disjoint, we can operationally distinguish between them with a success probability strictly larger than $\frac{1}{2}$, and any bipartite pure state with full Schmidt rank can be used as input.

The really interesting thing here is that the equality of $\widetilde{D}(\rC_0,\rC_1)=D(\rC_0,\rC_1)$ does hold. The key to this is the application of Sion's minimax theorem and the following semi-definite programming characterization of the diamond norm recently discovered by Watrous \cite{note1}.
\begin{lemma}\label{propWatrous}(Watrous \cite{Watrous09}) For any super-operator $\Phi=\Phi_0-\Phi_1$ such that $\Phi_0$ and $\Phi_1$ are quantum channels in $\rT(\cH,\cK)$, we have the following
$$||\Phi||_\diamond=\max 2\tr \rho_\Phi X, ~~X\leq I\otimes \rho, \tr\rho=1, \rho\geq 0, X\geq 0,$$
where $\rho_\Phi=(\Phi\otimes I_{\cH'})(\op{\alpha}{\alpha})$ is the Choi operator of $\Phi$, $\ket{\alpha}=\sum_{k=1}^d \ket{k}\otimes\ket{k}=\sqrt{d}\ket{\Omega}$ is the unnormalized maximally entangled state over $\cH\otimes\cH'$, and $\cH'$ is an isomorphic copy of $\cH$.
\end{lemma}

Now we can summarize the relation between $\widetilde{D}(\rC_0,\rC_1)$ and $D(\rC_0,\rC_1)$ as follows:
\begin{theorem}\label{DisTildeD}
Let $\rC_0$ and $\rC_1$ be two compact convex sets of quantum channels in $\rT(\cH,\cK)$. Then
$$\widetilde{D}(\rC_0,\rC_1)=D(\rC_0,\rC_1).$$
\end{theorem}

{\bf Proof:} Let us first denote $$\rC=\rC_0-\rC_1=\{\Phi_0-\Phi_1: \Phi_0\in \rC_0, \Phi_1\in \rC_1\}.$$ Then $\rC$ is a compact convex set, and completely determines $\widetilde{D}(\rC_0,\rC_1)$ and $D(\rC_0,\rC_1)$. We also write
$$R=\{(X,\rho): 0\leq X\leq I\otimes \rho, \rho\geq 0, \tr \rho=1\}.$$ Clearly, $R$ is also a compact convex set.

By Lemma \ref{propWatrous}, we can rewrite
$$D(\rC_0,\rC_1)=\min_{\Phi\in \rC} \max_{(X,\rho)\in R} 2\tr\rho_\Phi X.$$
Noticing that both $\rC$ and $R$ are compact convex sets, and the objective function $2\tr (\rho_\Phi X)$ is linear both in $\Phi$ and $(X,\rho)$, by Sion's minimax theorem we can exchange the order of ``max" and ``min" as follows:
$$D(\rC_0,\rC_1)=\max_{(X,\rho)\in R} \min_{\Phi\in \rC} 2\tr \rho_\Phi X.$$

Now we proceed to prove $\widetilde{D}(\rC_0,\rC_1)=D(\rC_0,\rC_1)$. We only need to show $\widetilde{D}(\rC_0,\rC_1)\geq D(\rC_0,\rC_1)$ as the opposite direction is obvious according the definitions. By the above equation and Eq. (\ref{tildeD}), it suffices to show that for any $(X,\rho)\in R$ there is a density operator $\sigma\in \rL(\cH\otimes \cH')$ such that
$$||(\Phi\otimes I)(\sigma)||_{1}\geq \tr \rho_{\Phi}X,\forall \Phi\in \rC.$$
Indeed, we can choose $\sigma=\op{u}{u}$ to be the following bipartite pure state
$$\ket{u}=(I\otimes A)\ket{\alpha} ~{\rm and}~A^\dagger A=\rho,$$
where $\ket{\alpha}$ is again the unnormalized maximally entangled state over $\cH\otimes \cH'$.

Note that we have the following well-known fact about the trace norm:
$$||Y||_1=\max_{0\leq P\leq I}2\tr PY,$$
where $Y$ is any traceless ($\tr Y=0$) Hermitian operator. Applying the above fact to $(\Phi\otimes I)(\sigma)$, we have
$$||(\Phi\otimes I)(\sigma)||_{1}=\max_P 2\tr P(I\otimes A)\rho_\Phi (I\otimes A^\dagger)=\max_{Q}2\tr \rho_\Phi Q,$$
where $0\leq P\leq I_{\cH\otimes \cH'}$ and $Q=(I\otimes A^\dagger)P(I\otimes A)$. Noticing that $0\leq X\leq I\otimes \rho=I\otimes A^\dagger A$, we can easily find $0\leq P'\leq I_{\cH\otimes \cH'}$ such that $X=Q'=(I\otimes A^\dagger)P'(I\otimes A)$ \cite{note2}. Thus we have
$$\max_{Q}2\tr \rho_\Phi Q\geq 2\tr \rho_\Phi Q'=2\tr \rho_\Phi X,$$
which completes the proof.\hfill $\square$

{\bf Remarks:} After we finished the above proof, we were informed by Gutoski that in a recent work he generalized the results in Ref. \cite{GW04} to the discrimination of two compact convex sets of quantum strategies, and obtained the results for the case of quantum channels as an immediate corollary \cite{Gutoski08}. It is interesting to note that his main proof technique is a separation theorem of compact convex sets from convex analysis, quite similar to that in Ref. \cite{GW04}. Instead, here we employ a different method by using Sion's minimax theorem and semi-definite programming characterization of diamond norm, in a similar spirit of Ref. \cite{Jain05}. Hopefully, our proof may provide some new insight into this problem. Gutoski's paper, however, contains many other interesting results about the trace norms.

It is also worth noting that with minor changes the same technique in the above proof can be used to derive Lemma \ref{propWatrous}, as first shown by Watrous in Ref. \cite{Watrous09}.

All the above discussions are applicable to the case of  $\rC_0=\{\Phi\}$ and $\rC_1={\rm Conv}(\bU(\cH))$. An interesting fact is that without auxiliary systems, we cannot operationally distinguish between a unital quantum channel $\Phi$ and ${\rm Conv}(\bU(\cH))$ even when the former is not contained in the latter. To see this, let $\rho\in \rL(\cH)$ be any density operator. Since $\Phi$ is a unital quantum channel, it is also a doubly stochastic map. Thus we have the majorization relation $\Phi(\rho)\prec\rho$ \cite{note3}. By another Theorem of Uhlmann \cite{Uhl77}, we know there exist a probability distribution $\{p_k\}$ and a set of unitary operators $\{U_k\}$ such that
$$\Phi(\rho)=\sum_k p_k U_k\rho U_k^\dagger.$$
So $\widetilde{D}(\Phi,{\rm Conv}(\bU(\cH));\rho)=0$ for any density operator $\rho$ from $\rL(\cH)$.
On the other hand, we have $D(\Phi,{\rm Conv}(\bU(\cH)))>0$ even when the input can only be chosen from $\rL(\cH)$.  This indicates that $D$ and $\widetilde{D}$ are quite different when we do not use auxiliary systems.

\section{a Lower bound for the distance between a quantum channel and the convex hull of unitary channels}

It is generally difficult to decide whether a given unital quantum channel $\Phi$ is a mixture of unitary channels or not. One simple sufficient condition is that the Kraus operator space $K(\Phi)$ does not contain any unitary operator, i.e., $K(\Phi)\cap \rU(\cH)=\emptyset$. (Note that the Kraus operator space $K(\Phi)={\rm span}\{E_k\}$ for a quantum channel $\Phi=\sum_k E_k\cdot E_k^\dagger$). If this is the case, we can actually obtain an analytical lower bound for the distance between $\Phi$ and ${\rm Conv}(\bU(\cH))$.

\begin{lemma}\label{keylemma}
For any quantum channel $\Phi\in \rT(\cH_d)$ such that $K(\Phi)\cap \rU(\cH_d)=\emptyset$, we have
$$D(\Phi, {\rm Conv}(\bU(\cH_d)))\geq \min_{L\in K(\Phi)}\frac{\tr (|L|-I_d)^2}{d}=C_\Phi>0.$$
\end{lemma}

{\bf Proof:} Let $\Psi=\sum_{k=1}^{N} p_k \cU_k$ with $\{p_k\}$ a finite probability distribution and $\cU_k\in \bU(\cH_d)$. We need to show that
$$D(\Phi,\Psi)=D(\Phi, \sum_k p_k \cU_k)\geq C_\Phi.$$
Note that
\begin{align*}
D(\Phi,\sum_k p_k \cU_k)=& D(\Phi\otimes I_{\rL(\cZ_d)},\sum_k p_k \cU_k\otimes I_{\rL(\cZ_d)})\nonumber \\
\geq & D(\Phi\otimes I_{\rL(\cZ_d)}(\Omega),\sum_k p_k \cU_k\otimes I_{\rL(\cZ_d)}(\Omega)),\nonumber
\end{align*}
where $\ket{\Omega}=1/\sqrt{d}\sum_{k=1}^d \ket{k}\ket{k}$ is a maximally entangled state on $\cH_d\otimes \cZ_d$.
Now applying the inequality $D(\rho,\sigma)\geq 2(1-F(\rho,\sigma))$, we have
\begin{align*}
D(\Phi,\Psi)\geq& 2(1-F(\Phi\otimes I_{\rL(\cZ_d)}(\Omega),\sum_k p_k \cU_k\otimes I_{\rL(\cZ_d)}(\Omega)))\nonumber \\
=& 2(1-\max_{\psi}|\ip{\psi}{\phi}|),\nonumber
\end{align*}
where  $\ket{\psi}=\sum_k \sqrt{q_k} \ket{\psi_k}\ket{k_\cK}$ ranges over all purifications of $\Phi\otimes I_{\rL(\cZ_d)}(\Omega)$, $\ket{\phi}$ is a fixed purification of $\sum_k p_k \cU_k\otimes I_{\rL(\cZ_d)}(\Omega)$ given by
$$\ket{\phi}=\sum_k \sqrt{p_k}(U_k\otimes I_{\cZ_d})\ket{\Omega}\otimes \ket{k_\cK},$$ $\{\ket{k_\cK}\}$ is a fixed orthonormal basis for an auxiliary system $\cK$, $\{q_k\}$ is a probability distribution, and $\ket{\psi_k}$ are unit vectors in $\cH_d\otimes \cZ_d$. An important observation here is that $\ket{\psi_k}$ is in the support of $\Phi\otimes I_{\rL(\cZ_d)}(\Omega)$ which is spanned by a set of vectors of the form $(E_j\otimes I_{\cZ_d})\ket{\Omega}$, where we assume that $\Phi=\sum_j E_j\cdot E_j^\dagger$. Hence
$$\ket{\psi_k}=\sum_j \lambda_j (E_j\otimes I_{\cZ_d})\ket{\Omega}$$
for some complex numbers $\lambda_j$, from which we readily deduce that
$$\ket{\psi_k}=(L_k\otimes I_{\cZ_d})\ket{\Omega},$$
where
$$L_k=\sum_j \lambda_j E_j\in K(\Phi).$$
Since $\ket{\psi_k}$ are unit vectors,  $\tr |L_k|^2=\tr(L_k^\dagger L_k)=d$. Thus we have
\begin{align*}
D(\Phi,\Psi)\geq&2(1-\max_{L_k, q_k}|\sum_k \sqrt{p_kq_k} \tr(L_k^\dagger U_k)/d|\})\nonumber\\
\geq & 2(1-\max \{\tr(L_k^\dagger U_k)/d: L_k\in K(\Phi)\})\nonumber \\
\geq & 2(1-\max\{\frac{1}{d}\tr(L^\dagger U):\tr|L|^2=d, U\in \rU(\cH_d)\})\nonumber\\
= &2(1-\frac{1}{d} \max \{\tr |L|: \tr |L|^2=d, L\in K(\Phi)\})\nonumber\\
= &\min \{\frac{1}{d}{\tr(|L|-I_{\cH_d})^2}:\tr |L|^2=d, L \in K(\Phi)\}\nonumber\\
\geq & \inf\{\frac{1}{d}{\tr(|L|-I_{\cH_d})^2}:L\in K(\Phi)\}.\nonumber
\end{align*}

In the last step we have to use ``$\inf$" instead of ``$\min$" as the domain of $L$ has been broadened from a compact set $\{L\in K(\Phi): \tr|L|^2=d\}$ to an unbounded set $K(\Phi)$. To finish the proof, we need to show that ``inf" in the last line can be replaced by ``min". First, notice that the RHS of the above equation is less than $2$, and
$$\frac{\tr (|L|-I_{\cH_d})^2}{d}\geq \frac{(\tr|L|-d)^2}{d^2}.$$
If $\tr|L|\geq (\sqrt{2}+1)d$ then the right hand side (RHS) of the above equation is greater than $2$. Thus
$$
\inf_{L\in K(\Phi)}\frac{\tr (|L|-I_{\cH_d})^2}{d}=\min_{\tr|L|\leq (1+\sqrt{2})d}\frac{\tr (|L|-I_{\cH_d})^2}{d}.
$$
As a final remark, we need show that $C_\Phi>0$ under the assumption $K(\Phi)\cap \rU(\cH_d)=\emptyset$. Otherwise, $C_\Phi=0$ implies that there is some $\widetilde{L}\in K(\Phi)$ such that $|\widetilde{L}|=I_{\cH_d}$. In other words, $\widetilde{L}$ is unitary, which is a contradiction.\hfill $\square$

\begin{theorem}
Let $\Phi\in \rT(\cH_d)$ be a quantum channel, and let $\Psi\in \rS(\cH_m)$ be any Schur channel. Then $$D(\Psi\otimes\Phi,{\rm Conv}(\bU(\cH_m\otimes \cH_d)))\geq C_\Phi.$$
\end{theorem}

{\bf Proof:} The key observation here is that under the assumption $\Psi$ is with diagonal Kraus operators. Thus any $L\in K({\Psi\otimes \Phi})$ can be decomposed as $$L=\oplus_{k=1}^m L_k,~~L_k\in K(\Phi).$$
Suppose now that $\widetilde{L}=\oplus_{k=1}^m \widetilde{L}_k$ achieves the minimum in $C_{\Psi\otimes \Phi}$. We have
\begin{align*}
C_{\Psi\otimes \Phi}=&\frac{1}{md}{\tr(|\widetilde{L}|-I_{\cH_m\otimes\cH_d})^2}\nonumber\\
=&\frac{1}{md}{\tr(\oplus_{k=1}^m (|\widetilde{L}_k|-I_{\cH_d}))^2}\nonumber\\
=&\frac{1}{md}{\tr(\oplus_{k=1}^m (|\widetilde{L}_k|-I_{\cH_d})^2)}\nonumber\\
=&\frac{1}{md}{\sum_{k=1}^m \tr (|\widetilde{L}_k|-I_{\cH_d})^2}\nonumber\\
\geq& C_\Phi,\nonumber
\end{align*}
where we have employed the fact that
$$\tr(|\widetilde{L}_k|-I_{\cH_d})^2\geq d C_\Phi, \forall~1\leq k\leq m.$$
Now the desired result follows from Lemma \ref{keylemma} directly. \hfill $\square$

As a direct corollary, we have the following
\begin{corollary}\label{ex-aqbc}
For any Schur channel $\Phi\in \rS(\cH_d)$, if $K(\Phi)\cap \rU(\cH_d)=\emptyset$, then
$$D(\Phi^{\otimes n},{\rm Conv}(\bU(\cH_d^{\otimes n})))\geq C_\Phi>0,~\forall n\geq 1.$$
\end{corollary}

\section{Bounds on the distance between a Schur channel and the convex hull of unitary channels}
The condition that the Kraus operator space $K(\Phi)$ of $\Phi$ does not contain any unitary operator is a very strong constraint. In most cases we may have that $\Phi$ is not a mixture of unitary channels but $K(\Phi)$ contains some unitary operator. Here we deal with this more general case but only for Schur channels. In this case we are able to show that up to a factor of $1/2$, any Schur channel can be approximated by a mixture of diagonal unitary channels.

Let us denote
$$\Lambda(\cH_d)=\rS(\cH_d)\cap {\rm Conv}(\bU(\cH_d)).$$
Intuitively, $\Lambda(\cH_d)$ (or simply $\Lambda_d$) is the set of Schur channels that are also mixtures of diagonal unitary channels. So any $\Psi\in \Lambda_d$ can be written into the form $\Psi=\sum_{k} p_k U_k\cdot U_k^\dagger$, where $U_k$ are $d\times d$ diagonal unitary matrices.
\begin{theorem}\label{schur-multiplier1}
For given Schur channel $\Phi\in \rS_d$, we have
\begin{equation}\label{schur-eq2}
\frac{1}{2}D(\Phi,\Lambda_d)\leq D(\Phi,{\rm Conv}(\bU_d))\leq D(\Phi,\Lambda_d).
\end{equation}
\end{theorem}

{\bf Proof:} The second inequality follows directly from $\Lambda_d\subset {\rm Conv}(\bU_d)$.  We will employ some standard arguments in operator algebras to prove the first inequality. Let $\Psi=\sum_k p_k \cU_k\in {\rm Conv}(\bU_d)$ such that
$$D(\Phi, \Psi)=D(\Phi,{\rm Conv}(\bU_d))=\delta.$$
We only need to prove that
$$D(\Phi,\Lambda_d)\leq 2\delta.$$
For any two diagonal unitary matrices $U$ and $V$, let us introduce a map $\rJ^{U,V}: \rT(\cH_d)\rightarrow \rT(\cH_d)$ as follows:
$$\rJ^{U,V}(\Phi)=U^\dagger\Phi(U\cdot V)V^\dagger.$$
It is obvious that $\rJ^{U,V}$ is an isometry over $\rT(\cH_d)$ in the following sense:
\begin{equation}\label{juv}
D(\rJ^{U,V}(\Phi_1),\rJ^{U,V}(\Phi_2))=D(\Phi_1,\Phi_2)
\end{equation}
for any $\Phi_1,\Phi_2\in \rT(\cH_d)$.

Now we can further introduce a map $\rJ: \rT(\cH_d)\rightarrow \rT(\cH_d)$ such that
$$\rJ(\Phi)=\int_{\widehat{\rU}_d}\int_{\widehat{\rU}_d}\rJ^{U,V}(\Phi) dU dV,$$
where both $dU$ and $dV$ are Haar measures over the diagonal unitary group $\widehat{\rU}_d$.
The map $\rJ$ satisfies the following properties:\\
i). $\rJ$ is a contraction in the sense
$$D(\rJ(\Phi_1),\rJ(\Phi_2))\leq D(\Phi_1,\Phi_2), \forall \Phi_1,\Phi_2\in \rT(\cH_d),$$
which is a simple consequence of the convexity of the diamond norm and Eq. (\ref{juv}).

ii). $\rJ(\Phi)=\Phi$ for any Schur multiplier $\Phi\in \rT(\cH_d)$. This is true simply due to the following observation
$$\rJ^{U,V}(\Phi)=\Phi, $$
where $U,V\in \widehat{\rU}_d$ are diagonal unitary matrices.

iii). $\rJ(\Phi)$ is a Schur multiplier for any $\Phi\in \rT(\cH_d)$. In particular, $\rJ(\Phi)$ is CP whenever $\Phi$ is CP.  To see that, by a direct calculation, we find that for $\Phi=\sum_k E_k\cdot F_k^\dagger$,
$$\rJ(\Phi)=\sum E_k'\cdot F_k'^\dagger, ~E_k'={\rm diag}(E_k), F_k'={\rm diag}(F_k).$$ Clearly, $\rJ(\Phi)$ is a Schur multiplier. When $\Phi$ is CP, we can choose
$$E_k'=F_k'={\rm diag}(E_k)={\rm diag}(F_k),$$ thus $\rJ(\Phi)$ is CP.

Now we can compute that
$$\Psi'=\rJ(\Psi)=\sum_k p_k \rJ(\cU_k)=\sum_k p_k A_k\cdot A_k^\dagger,$$ where $A_k={\rm diag}(U_k)$.
We have
\begin{equation}\label{triangle}
D(\Phi,\Lambda_d)\leq D(\Phi, \Psi')+D(\Psi',\Lambda_d).
\end{equation}
The first term in the RHS of Eq. (\ref{triangle}) satisfies
$$D(\Phi, \Psi')=D(\rJ(\Phi),\rJ(\Psi))\leq D(\Phi,\Psi)=\delta,$$
where we have employed the contraction property of $\rJ$, item i) above.

It remains to show that the second term in the RHS of Eq. (\ref{triangle}) fulfills
$$D(\Psi',\Lambda_d)\leq \delta.$$
Our strategy is to choose $\Psi''\in \rT(\cH_d)$ such that
\begin{equation}\label{immediate-state}
D(\Psi',\Psi'')\leq \delta,~\Psi''\in \Lambda_d
\end{equation}
Then
$$D(\Psi',\Lambda_d)\leq D(\Psi',\Psi'')\leq \delta.$$
The rest of the proof devotes to finding such $\Psi''$. Notice that each Kraus operator $A_k$ of $\Psi'$ is a diagonal contraction. Applying a well-known fact in linear algebra, we can choose two diagonal unitary matrices $V_k$, $W_k$ such that
$$A_k=\frac{1}{2}(V_k+W_k).$$
Now define $$\Psi''=\sum_k \frac{p_k}{2}({V_k\cdot V_k^\dagger+W_k\cdot W_k^\dagger}).$$ Clearly $\Psi''\in \Lambda_d$. We will show that $\Psi''$ satisfies Eq. (\ref{immediate-state}). First, we find that $\Psi''-\Psi'$ is a CP map. By a direct calculation, we have
$$\Psi''-\Psi'=\sum_k \frac{p_k}{4}(W_k-V_k)\cdot(W_k-V_k)^\dagger.$$
Thus
$$(\Psi''-\Psi')^\dagger=\Psi''^\dagger-\Psi'^\dagger$$ is also a CP map.

Now employing essentially the same techniques first introduced by Haagerup and Musat in \cite{HM2011b}, we have
\begin{align*}
||\Psi''-\Psi'||_\diamondsuit =&||\Psi''^\dagger-\Psi'^\dagger||_{\rm cb}\nonumber\\
=&||\Psi''^\dagger-\Psi'^\dagger||_{\rm \infty}\nonumber\\
=&||(\Psi''^\dagger-\Psi'^\dagger)(I_{\cH_d})||_{\infty}\nonumber\\
=&||I_{\cH_d}-\Psi'^\dagger (I_{\cH_d})||_{\infty}\nonumber\\
=&||\Phi^\dagger(I_{\cH_d})-\Psi'^\dagger (I_{\cH_d})||_{\infty}\nonumber\\
\leq&||\Phi^\dagger-\Psi'^\dagger||_\infty\nonumber\\
=&||\Phi-\Psi'||_1\nonumber\\
\leq&||\Phi-\Psi'||_\diamond.\nonumber
\end{align*}
That means
$$D(\Psi',\Psi'')\leq D(\Phi,\Psi')\leq \delta.$$\hfill $\square$

It seems quite likely that in Eq. (\ref{schur-eq2}) the first inequality should be strict and the second one should be an equality. However, this is still an unsettled issue.

\begin{theorem} \label{schur-multiplier2}
For given Schur channel $\Phi\in\rS_d$ and arbitrary $\Psi\in \rS_m$, we have
\begin{equation}\label{diagonal1}
D(\Psi\otimes \Phi, \Lambda_{m\otimes d})\geq D(\Phi,\Lambda_d), \forall \Psi\in \rS_m.
\end{equation}
Here $m\otimes d$ is a shorthand for $\cH_m\otimes \cH_d$.
\end{theorem}

{\bf Proof:} To show Eq. (\ref{diagonal1}), we first choose $\Psi'=\sum_k p_k U_k\cdot U_k^\dagger \in \Lambda_{m\otimes d}$ such that
\begin{equation}\label{diagonal2}
D(\Psi\otimes\Phi, \Lambda_{m\otimes d})=D(\Psi\otimes\Phi, \Psi').
\end{equation}
Notice that any diagonal unitary matrix $U_k$ can be written into the following form:
$$U_k=\sum_{j=1}^m \op{j}{j}\otimes U_{j}^{(k)},$$
where $U_{j}^{(k)}\in \widehat{\rU}_d$ are all diagonal unitary matrices.
That implies
$$\Psi''=\braket{1}{\Psi}{1}=\sum_{k}p_k \cU_{1}^{(k)}\in \Lambda_d,$$
where $\cU_{1}^{(k)}$ is the unitary channel corresponding to $U_{1}^{(k)}$. Intuitively, the compressed version $\Psi''$ of $\Psi'$ remains a mixture of diagonal unitary channels.
Now we have
\begin{align*}
D(\Psi\otimes \Phi,\Psi')\geq & \sup_{||X||_1\leq 1}||(\Psi\otimes \Phi-\Psi')(\op{1}{1}\otimes X)||_1\\
=& \sup_{||X||_1\leq 1}||\op{1}{1}\otimes (\Phi-\Psi'')(X)||_1\\
=& ||\Phi-\Psi''||_1\\
=& ||\Phi-\Psi''||_\diamondsuit\\
\geq & D(\Phi,\Lambda_d),
\end{align*}
where we have used the fact that $\Psi(\op{1}{1})=\op{1}{1}$ and $\Psi''=\braket{1}{\Psi'}{1}\in \Lambda_d$.

Combining the above equation with Eq. (\ref{diagonal2}), we have proven Eq. (\ref{diagonal1}). \hfill $\square$

A somewhat interesting fact is that the above two results together can be used to derive some results first obtained by Haagerup and Musat in Ref. \cite{HM2011b}, which are applicable to the AQBC.
\begin{theorem}{(Haagerup and Musat \cite{HM2011b})}\label{hm-theorem1}
For given Schur channel $\Phi\in\rS_d$ and arbitrary $\Psi\in \rS_m$, we have
\begin{equation}\label{diagonal}
D(\Psi\otimes \Phi, {\rm Conv}(\bU_{m\otimes d}))\geq \frac{1}{2}D(\Phi,{\rm Conv}(\bU_d)).
\end{equation}
\end{theorem}

{\bf Proof:} Actually Eq. (\ref{diagonal}) is a quite straightforward application of the above two theorems. First notice that $\Psi\otimes \Phi\in \rS_{m\otimes d}$. Applying Theorem \ref{schur-multiplier1} to $\Psi\otimes \Phi$, we have
$$D(\Psi\otimes \Phi, {\rm Conv}(\bU_{m\otimes d}))\geq \frac{1}{2} D(\Psi\otimes \Phi, \Lambda_{m\otimes d}).$$
On the other hand, it is obvious that
$$D(\Phi,\Lambda_d)\geq D(\Phi, {\rm Conv}(\bU_d)).$$
Thus the left thing is to show
$$D(\Psi\otimes \Phi, \Lambda_{m\otimes d})\geq D(\Phi,\Lambda_d),$$
and this is exactly the content of Theorem \ref{schur-multiplier2}. \hfill $\square$

\begin{corollary}{(Haagerup and Musat \cite{HM2011b})} \label{ex-aqbc2}
Let $\Phi\in \rS_d$ be a Schur channel that does not satisfy the quantum Birkhoff property, that is, $\Phi\not\in {\rm Conv}(\bU_d)$. Then $\Phi$ does not satisfy the asymptotic quantum Birkhoff property, and
$$D(\Phi^{\otimes n}, {\rm Conv}(\bU_{d^{\otimes n}}))\geq \frac{1}{2}D(\Phi, {\rm Conv}(\bU_d)).$$
\end{corollary}

\section{Some explicit counterexamples to the Asymptotic Quantum Birkhoff Conjecture}

Our results in Section IV enable us to construct counterexamples to AQBC easily. Our basic strategy is to construct Schur channel $\Phi$ satisfying $K(\Phi)\cap \rU(\cH)=\emptyset$. Then the statement that $\Phi$ is a counterexample to AQBC follows directly from Corollary \ref{ex-aqbc}.
\begin{example}\upshape
Our first example is chosen from Ref. \cite{LS93} (Section 4.3). $\Phi=E_1\cdot E_1^\dagger+E_2\cdot E_2^\dagger$, where
$$E_1={\rm Diag}(1,0,\frac{1}{\sqrt{2}},\frac{1}{\sqrt{2}}),~E_2={\rm Diag}(0,1,\frac{1}{\sqrt{2}},-\frac{i}{\sqrt{2}}).$$
Clearly none of $E_1$ and $E_2$ is unitary. We now show that there is no unitary in $K(\Phi)$. By contradiction, assume that for some complex numbers $\lambda$ and $\mu$ we have that $\lambda E_1+\mu E_2$ is unitary. Then
$$(\lambda E_1+\mu E_2)^\dagger (\lambda E_1+\mu E_2)=I_4,$$ from which we obtain
$$|\lambda|^2=1, |\mu|^2=1, \frac{1}{2}(|\lambda|^2+|\mu|^2)=1, \frac{1}{2}(|\lambda|^2-|\mu|^2)=1.$$
Clearly, there is no $\lambda$ and $\mu$ satisfying all the above equations. Thus we have $K(\Phi)\cap \rU(\cH_4)=\emptyset$. It follows from Corollary \ref{ex-aqbc} that $\Phi$ is a counterexample to AQBC. One can readily verify that the set $\{E_1^\dagger E_1, E_1^\dagger E_2, E_2^\dagger E_1, E_2^\dagger E_2\}$ is linearly independent. Thus it follows from Corollary 2.3 of Ref. \cite{HM2011a} that $\Phi$ is a non-factorizable map. \hfill $\square$
\end{example}

\begin{example}\upshape
Our second example is taken from Ref. \cite{HM2011a} (Example 3.3). $\Phi=\sum_{k=1}^3 E_k\cdot E_k^\dagger$, where
$$E_1={\rm Diag}(1,\frac{1}{\sqrt{5}}I_5), E_2={\rm Diag}(0, \sqrt{\frac{2}{5}}Z_5), E_3=E_2^\dagger,$$
where $Z_5={\rm Diag}(1,\frac{2\pi i}{5},\frac{4\pi i}{5},\frac{6\pi i}{5},\frac{8\pi i}{5})$ satisfying $Z_5^5=I_5$. In the following we directly write $I$ and $Z$ for $I_5$ and $Z_5$, respectively.

As shown in \cite{HM2011a}, one can choose a set of Hermitian Kraus operators $F_1, F_2,F_3$ such that
$$F_1=E_1,~F_2=\frac{1}{2}(E_2+E_3),~F_3=\frac{1}{2i}(E_2-E_3).$$
It is easy to see that $\Phi=\sum_{k=1}^3 F_k\cdot F_k^\dagger$. By Corollary 2.5 of Ref. \cite{HM2011a}, $\Phi$ is  a factorizable map.

Now we show that $K(\Phi)\cap \rU(\cH_6)=\emptyset$. Again by contradiction, assume there are complex numbers $\lambda_1,\lambda_2,\lambda_3$ such that $\lambda_1E_1+\lambda_2 E_2+\lambda_3E_3$ is a unitary. In other words, ${\rm Diag}(\lambda_1,\sqrt{\frac{1}{5}}\lambda_1 I+\sqrt{\frac{2}{5}}\lambda_2 Z+\sqrt{\frac{2}{5}} \lambda_3 Z^{-1})$ is a unitary. This is equivalent to
$$|\lambda_1|^2=1,~|\lambda_1I+\sqrt{2}\lambda_2 Z+\sqrt{2}\lambda_3 Z^{-1}|=\sqrt{5}I.$$
For simplicity, we may assume $\lambda_1=1$, $a=\sqrt{2}\lambda_2$, and $b=\sqrt{2}\lambda_3$. Then we can rewrite the above equation as follows:
$$(|a|^2+|b|^2-4)I+(a+b^*)Z+(a^*+b)Z^{-1}+ab^* Z^2+ab^* Z^{-2}=0.$$
Employing the fact that $\{I,Z,Z^{-1}, Z^2, Z^{-2}\}$ are linearly independent, we have
$$|a|^2+|b|^2=4, a+b^*=0,ab^*=0.$$
Clearly, there are no $a$ and $b$ satisfying all the above equations.

Hence Corollary \ref{ex-aqbc} is applicable. This gives us a factorizable map which is also a counterexample to AQBC. This fact has been pointed out in the published version of Ref. \cite{HM2011a}, and was derived by the result in Ref. \cite{HM2011b}. \hfill $\square$
\end{example}

As a matter of fact, all counterexamples to AQBC presented above are simply the counterexamples to QBC. It would be quite interesting to know for what kind of unital channels $\Phi$ these two properties are different, i.e., $\Phi$ is a counterexample to QBC, but fulfills AQBC. A systematic way to construct unital channels that violate QBC has been proposed by Bravyi and Smolin using the idea of unextendible maximally entangled bases \cite{BS11}. All unital channels $\Phi$ constructed in this way will automatically satisfy the condition $K(\Phi)\cap \rU(\cH)=\emptyset$.
However, it remains a formidable task to verify whether these unital channels fulfill or violate the AQBC. A preliminary step towards this goal is to invent some tractable upper bounds for the distance between a unital channel and the convex hull of unitary channels.

\section{A connection to Grothendieck's inequality}
It is well known that Grothendieck's inequality (GI) in the metric theory of tensor products is closely related to Bell's inequality in quantum information theory \cite{Tsirelson87}. As an interesting application of the results in Sections IV and V, we explain  here that GI has intimate links with AQBC too. Let us first recall the equivalent formulation of GI in terms of Schur multipliers. Let
$$\mathbb S_d=\{\Phi_S: S\in \rL(\cH_d), \|\Phi_S\|_1\le1\}.$$
Namely, $\mathbb S_d$ is the unit ball of the space of all Schur multipliers on $\rL(\cH_d)$ with respect to the trace norm or any of the three other norms considered before (see Proposition \ref{schur-multiplier3}). It is well known that $\Phi_S\in\mathbb S_d$ if and only if there exists a Hilbert space $\cK$ and vectors $\ket{\xi_1},..., \ket{\xi_d}, \ket{\eta_1}, ..., \ket{\eta_d}$ in the unit ball of $\cK$ such that
 \beq\label{schur}
 s_{kj}=\ip{\xi_k}{\eta_j}.
 \eeq
Here we have assumed that $S=[s_{kj}]$ with respect to a fixed orthonormal basis of $\cH_d$. Note that $\cK$ can be chosen to be finite dimensional.  This representation of $\Phi_S$ is to be compared with i) of Proposition \ref{schur0}. Indeed, developing $\xi_k$ and $\eta_j$ in an orthonormal basis of $\cK$, we recover  the representation of $\Phi_S$ given by Proposition \ref{schur0}.

The case where $\dim \cK=1$ is of particular interest. The corresponding set of  Schur multipliers is denoted by $\mathbb D_d$, that is,
$\mathbb D_d$ is the set  of all Schur multipliers $\Phi_S$ of the form
 \beq\label{schur diag}
 s_{kj}=\a_k^*\b_j, \quad 1\le k, j\le d,
 \eeq
where $\a_k, \b_j$ are complex numbers such that
 $$\max_k|\a_k|\le 1, \quad \max_j|\b_j|\le 1.$$
It is clear that
 $${\rm Conv}(\mathbb D_d)\subseteq\mathbb S_d.$$
GI asserts that the converse inclusion also holds true up to a universal constant:

\medskip

\noindent{\bf Grothendieck's inequality.} \emph{There exists a universal constant $K$ such that for all $d\ge1$}
 \beq\label{gro0}
 \mathbb S_d\subseteq K{\rm Conv}(\mathbb D_d).
 \eeq

The smallest constant $K$ is called Grothendieck's constant, denoted by $K_G$. The exact value of $K_G$ is still unknown. But it is well known that $1<K_G\leq 1.4049$. (Note here all scalars are assumed to be complex numbers). What is important for us is the fact that $K_G>1$. We refer to Chapter 5 of Ref. \cite{Pisier01} and page 19 of Ref. \cite{Pisier11} for more information.

\medskip

It is easy to see that a Schur multiplier $\Phi_S$ is  positive if and only if we can choose $\xi_k$ and $\eta_k$ in \eqref{schur} such that $\xi_k=\eta_k$ for all $1\le k\le d$. Let $\mathbb S_d^+$ denote the positive part of $\mathbb S_d$. Accordingly, let $\mathbb D_d^+$ denote the positive part of $\mathbb D_d$. Namely, $\mathbb D_d^+$ is the set of all Schur multipliers of the form \eqref{schur diag} with $\a_k=\b_k$. We again have obviously
 $${\rm Conv}(\mathbb D_d^+)\subseteq\mathbb S_d^+.$$
Surprisingly, this time the converse inclusion does not hold up to  a universal constant. More precisely, let $K_d^+$ denote the least constant such that
   $$\mathbb S_d^+\subseteq K_d^+{\rm Conv}(\mathbb D_d^+).$$
Then we have the following result of Kashin and Szarek from Ref. \cite{KS03} (see also the lemma on the page 17 of Ref. \cite{Pisier11}).

 \begin{proposition}\label{KS}
There exist two positive constants $\a$ and $\b$ such that $\a \log d\le K_d^+\le\b \log d$ for all $d>1$.
\end{proposition}

\medskip

The Schur multipliers we are interested in are  Schur channels (unital positive Schur multipliers). Recall that the set of  all Schur channels on $\rL(\cH_d)$ have been denoted by $S_d$ in the previous sections. This set could be also denoted by $\mathbb S_{d, 1}^{+}$ in the current notational system ($1$ being for ``unital").  Accordingly, $\Lambda_d$ is the subset of ${\rm Conv}(\mathbb D_d^+)$  consisting of Schur channels.   As shown in sections IV and V, to disprove AQBC is equivalent to showing that the obvious inclusion
 \beq\label{gro}
 \Lambda_d\subseteq S_{d}
 \eeq
is strict for some $d$. We now show that this inclusion is strict for large $d$ in the spirit of Proposition \ref{KS}. To this end let $K_{d, 1}^+$ denote the least constant $K$ such that
 $$S_{d}\subseteq K {\rm Conv}(\mathbb D_d^+)\,.$$
Then inclusion \eqref{gro} is strict if and only if $K_{d, 1}^+>1$.
 \begin{proposition} $K_{d, 1}^+ =K_{d}^+$ for all $d$.
  \end{proposition}

 {\bf Proof:} It is clear  that $K_{d,1}^+\le K_{d}^+$. To prove the converse inequality, let $\Phi_S\in\mathbb S_d^+$. Then there exist a Hilbert space $\cK$ and vectors $\xi_1, ..., \xi_d$ in the unit ball of $\cK$ such that
 $$s_{kj}=\ip{\xi_k}{\xi_j}.$$
Without loss of generality, we can assume that $\xi_k\neq0$ for all $k$. Let
$$\eta_k=\frac{\xi_k}{\|\xi_k\|}\quad\mbox{and} \quad t_{kj}=\ip{\eta_k}{\eta_j}.$$
Then $\Phi_T\in S_d\subseteq K_{d, 1}^+{\rm Conv}(\mathbb D_d^+)$, so there exist complex numbers $\a_{k,i}$ of modulus not greater than $1$ and positive numbers $\l_i$ such that
$$ t_{kj}=K_{d, 1}^+\sum_i\l_i\a_{k,i}^*\a_{j,i}\quad\mbox{and} \quad \sum_i\l_i=1.$$
It follows that
$$ s_{kj}=K_{d, 1}^+\sum_i\l_i(\|\xi_k\|\a_{k,i}^*)(\|\xi_j\|\a_{j,i})\in K_{d, 1}^+{\rm Conv}(\mathbb D_d^+).$$
We then deduce $K_{d}^+\le K_{d, 1}^+$.  \hfill $\square$

 Consequently,  $K_{d,1}^+\approx \log d$ as $d\to\8$. This implies that inclusion \eqref{gro} is strict for large $d$. Therefore, for any sufficiently large $d$, there exists a Schur channel $\Phi$ over $\rL(\cH_d)$ that is not a mixture  of unitary Schur channels. As direct consequences of the results in section V, such a Schur channel must violate AQBC.

By Ref. \cite{LS93}  the first integer $d$ for which $K_{d,1}^+>1$ is $d=4$. Thus  $K_{d}^+ =K_{d, 1}^+=1$ if and only if $d\le3$. This means that inclusion \eqref{gro} is an equality  for $d\le 3$ and becomes strict for $d\ge 4$.

On the other hand, denote $K_d$ the least constant $K$ in \eqref{gro0} for a fixed $d$.  Note that $K_G=\sup_dK_d$. It was proved independently by Davie and the third named author that $K_2=1$ (see the remark at the end of Chapter 5 of Ref. \cite{Pisier01}). It seems, however, that the first integer $d$ for which $K_d>1$ is still unknown. This problem is related to the characterization of the extreme points of $\mathbb S_d$. Indeed, $K_d>1$ is equivalent to the existence of extreme points $\Phi_S$ of $\mathbb S_d$ that are not of the form $S=[\a_k^*\b_j]$ for some complex numbers $\a_k$ and $\b_k$ with $|\a_k|=|\b_k|=1$ for all $k$. Thus such extreme points exist for large $d$. It would be interesting to characterize the extreme points of $\mathbb S_d$ in the spirit of Refs. \cite{Cho75} and \cite{LS93}.

\section*{Acknowledgements}
Part of this work was finished while R.D. and Q.X. were participating the quantum information theory program at the Mittag-Leffler Institute in the October of 2010, Sweden. The hospitality and the financial support of the organizers and institute were sincerely acknowledged. We especially thank M. Musat for carefully explaining their new results \cite{HM2011b} during the program, which has helped us to finish the proof of Theorem \ref{schur-multiplier1}. We were also indebted to an anonymous referee of QIP'2012 and J. Watrous for their helpful hints on the proof of Theorem \ref{DisTildeD}, and to G. Gutoski for informing us his relevant work, namely Ref. \cite{Gutoski08}, and for some interesting discussions during QIP'2012. R.D. was grateful to A. Winter for sharing his insight on this problem and for many delightful discussions, to M. B. Ruskai for her kind help during the program. N.Y. and R.D. were indebted to M. Ying for his constant support during this project.

This work was partly supported by the National Natural Science Foundation of China
(Grant Nos. 61179030 and 60621062), the Australian Research Council (Grant
Nos. DP110103473 and DP120103776), and Agence Nationale de Recherche (Grant No. 2011-BS01-008-01).


\begin{thebibliography}{99}
\bibitem{note2} This follows from a simple but useful fact in matrix analysis: For any $C^\dagger C\leq A^\dagger A$, there is $0\leq P\leq I$ such that $C^\dagger C=A^\dagger P A$.
\bibitem{note3} Let $x$ and $y$ be two real vectors with the same dimension. we say that $x$ is majorized by $y$, or $x\prec y$, if there is a doubly stochastic matrix $D$ such that $x=Dy$. For two Hermitian matrices $A$ and $B$, $A\prec B$ means $\lambda(A)\prec \lambda (B)$, where $\lambda(A)$ is the spectral vector of $A$, i.e., the real vector formed by all eigenvalues of $A$ (counting the mutilicity of eigenvalues).
\end{thebibliography}
\end{document}